\documentclass[onecolumn,aps,superscriptaddress,pre]{revtex4}

\usepackage{amsmath,amssymb,graphicx}
\usepackage{mathtools}

\newcommand{\Veff}{V_{\rm eff}}
\newcommand{\vin}{v_{\rm in}}
\newcommand{\vc}{v_{\rm c}}
\DeclareMathOperator{\sech}{sech}
\newcommand{\uK}{u_{\mathrm{K}}}
\newcommand{\uKbar}{u_{\bar{\mathrm{K}}}}
\newcommand{\uKK}{u_{\mathrm{KK}}}
\newcommand{\uKKbar}{u_{\mathrm{K}\bar{\mathrm{K}}}}

\newcommand{\cL}{\mathcal{L}}
\newcommand{\ws}{\omega_{\mathrm{s}}}

\begin{document}

\title{Four Decades of Kink Interactions in Nonlinear Klein-Gordon Models: \\
  A Crucial Typo, Recent Developments and the Challenges Ahead}

\author{P. G. Kevrekidis}
\email{kevrekid@math.umass.edu}
\affiliation{Department of Mathematics and Statistics, University of Massachusetts,
Amherst, Massachusetts 01003-4515 USA}

\author{R.H. Goodman}
\affiliation{Department of Mathematical Sciences,
New Jersey Institute of Technology,
University Heights, Newark, NJ 07102, USA}

\begin{abstract}
  The study of kink interactions in nonlinear
 Klein-Gordon models in $1+1$-dimensions
  has a time-honored history. Until a few years ago, it was arguably considered a fairly mature field whose main phenomenology was well understood both qualitatively and at least semi-quantitatively. This consensus was shattered when H. Weigel and his group established that the effective
  model that had allowed this detailed understanding contained  an all-important typo. Remarkably, they found that correcting this error wipes out both the quantitative and 
  qualitative agreement and, in fact, leads to additional problems.
  We summarize the history of the subject from the early studies, up to Weigel's work and reflect on where these recent developments
  leave our understanding (which, quantitatively, is close to square one!).
  Importantly, we stress a number of emerging additional directions
  that have arisen in higher-order power law models and speculate on the associated significant potential for future work.
\end{abstract}

\maketitle

\section{Preamble Teaser: A Mistake!}

Intriguingly, mathematics and science occasionally 
benefit significantly from a mistake that propels an area forward. Arguably,  the most famous such
example is the mistake in  Poincar{\'e}'s entry in
the competition to construct a convergent series solution to the three-body gravitational problem, sponsored by King Oscar II of Sweden
and Norway. After accepting the prize, Poincar{\'e} discovered a
significant problem with his calculations. This, in turn, led him
to discover the phenomena of sensitive dependence on initial conditions and chaos. Diacu and Holmes beautifully chronicle this story in~\cite{celestial} and we recommend it to anyone working in Dynamical Systems, as it describes, effectively, the genesis of the field.

Our exposition is centered around another mistake of this
kind. It, in fact, a appears far more innocuous-looking, as it only concerns
an apparent typographical error. Yet, it  has proved so detrimental that it
has set a seemingly mature and well-understood field in complete
disarray: much of the qualitative and semi-quantitative theory
can, remarkably, no longer be considered applicable (except
in a phenomenological way). But let us go back to the beginning.

\section{In the beginning, there was integrability\ldots\ }

In the 1970's , the discovery of the magical-seeming theory of completely integrable nonlinear waves revolutionized the study of nonlinear waves and led to a Steele Prize for its founders~\cite{Steele2006}. The centerpiece of this theory, the inverse
scattering transform (IST), has since been summarized in numerous
books~\cite{ablowitz,ablowitz2,drazin}. A major consequence of this  theory  is that the interaction of solitary waves is \emph{perfectly elastic} in integrable field theories, most notably in the many 1+1 dimensional examples. Solitary waves in such systems are called  solitons and emerge from collisions with their form and velocity unchanged, modulo
a so-called phase shift (a displacement from their undisturbed trajectory).
Relevant examples include nonlinear Klein-Gordon equations
such as the famous sine-Gordon (sG) equation~\cite{gibbon,oursg}
\begin{equation}
  u_{tt}=u_{xx}-\sin(u),
  \label{eq0}
  \end{equation}
which arises in models of superconducting Josephson junctions, coupled torsion
pendula, and surfaces of constant negative curvature (among many
other applications). Remarkably, similar behavior  arises in
the universal
nonlinear Schr{\"o}dinger equation~\cite{nls,abl3} used to model 
fluids and superfluids, optics and plasmas, and many other applications.
Note that $x$ and $t$ subscripts will be used hereafter to denote
space and time partial derivatives, while $u$ will be
used to denote the spatio-temporally dependent field.

\begin{figure}
    \centering
    \includegraphics[width=0.8\textwidth]{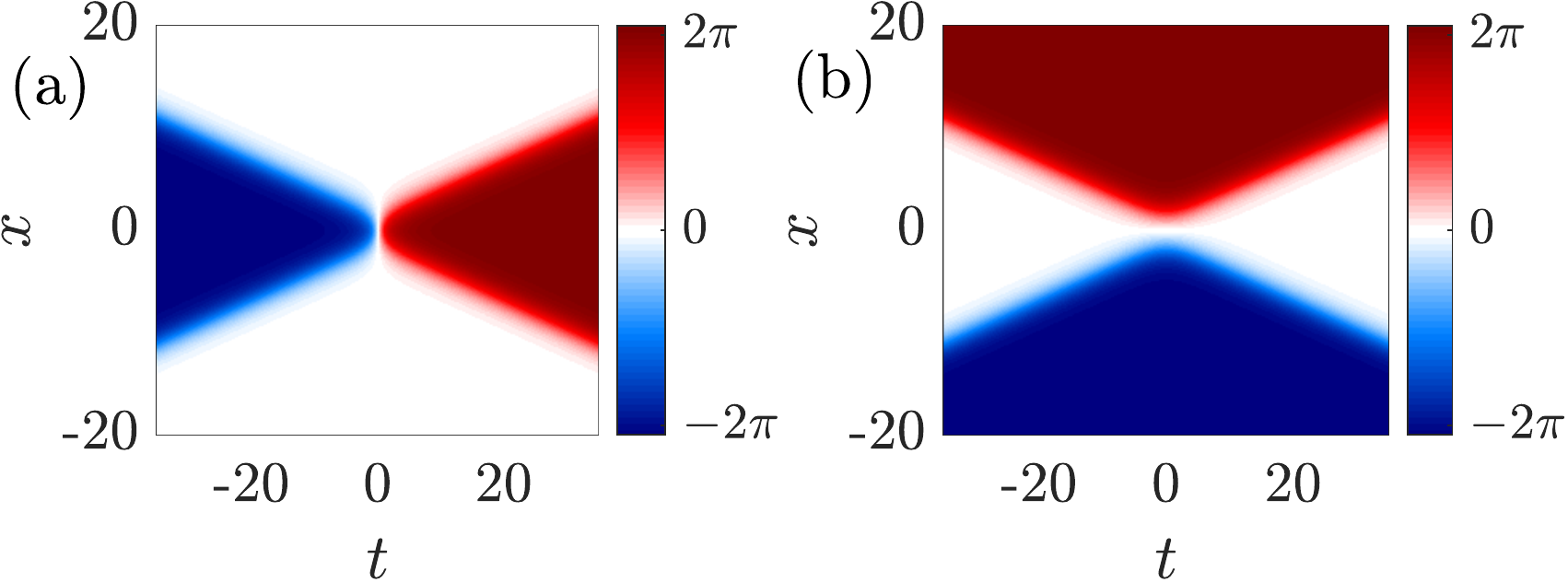}
    \caption{(a) The kink-antikink solution and (b) the kink-kink solution, both with $v=0.28$.}
    \label{fig:SG2kinks}
\end{figure}

Simple dynamical systems methods show that the sG model possesses single-soliton solutions in the form of kinks:
  \begin{equation}
  \uK(x,t)=4 \tan^{-1}\left(e^{\frac{x-x_0-v t}{\sqrt{1-v^2}}}\right), \, -1<v<1,
  \label{eq1}
  \end{equation}
  and antikinks $\uKbar(x,t)=\uK(-x,t)$. These two solutions are simply heteroclinic orbits connecting the spatially homogeneous solutions given by stable fixed points, and are
  centered initially at $x_0$, or $-x_0$ in the above antikink definition, and potentially
  traveling with speed $v$ as a result of  Lorentz invariance. The IST methods leveraged the  miracles of complete integrability
  (such as their surprising nonlinear superposition principles)
  to construct from simpler single-kink solutions, more elaborate ones
  such as kink-antikink and kink-kink solutions, in the form:
\begin{equation}
\uKKbar(x,t)=4 \tan^{-1}\left(\frac{\sinh{\frac{v t}{\sqrt{1-v^2}}}}
{v \cosh{\frac{x}{\sqrt{1-v^2}}}}\right), \quad
 \uKK(x,t)=4 \tan^{-1}\left(\frac{v \sinh{\frac{x}{\sqrt{1-v^2}}}}
{\cosh{\frac{v t}{\sqrt{1-v^2}}}}\right).
\label{eq2}
  \end{equation}
These are plotted in Fig.~\ref{fig:SG2kinks}
Although it may not appear obvious from the formulas, at long times these solutions aysmptotically separate, respectively, into the sum of a kink and anti-kink and the sum of two kinks.

Such exact solutions allow us infer that kinks and anti-kinks
attract each other dynamically, while two kinks repel, as shown by the following simple but entertaining calculation. As a single kink runs from $u=0$ to $u=2 \pi$, its center is considered to be located at $u=\pi$. Then, 
$\uKKbar=\pi$ leads to
\begin{equation}
  \tan{\left(\frac{u}{4}\right)}= \frac{\sinh{\frac{v t}{\sqrt{1-v^2}}}}
  {v \cosh{\frac{x}{\sqrt{1-v^2}}}} \Rightarrow
  v \cosh(\frac{x}{\sqrt{1-v^2}})=\sinh(\frac{v t}{\sqrt{1-v^2}}).
  \label{eq3}
\end{equation}
Assuming $v \ll 1$ (i.e., the Newtonian, rather than
relativistic regime) and working in the sector $x\gg 1$, $t\gg 1$, we may extract the equation for
the center located at $x$:
\begin{equation}
x= -\log(v) + v t -e^{-2 v t} \Rightarrow \ddot{x}=-4 e^{-2 x}, \quad \Veff(s)=-32 e^{- s}.
\label{eq3a}
\end{equation}
To define the kinetic and potential energy of kink-antikink attraction,  we have introduced the separation, $s=2 x$, and the notion of kink mass, 
\begin{equation}\label{kinkMass}
M=\int u_x^2 dx, 
\end{equation}
which in the sG case equals $8$. We can immediately infer the attractive nature of
the interaction from the sign of $V'(x)$. An equivalent calculation finds repulsion in the case of kink-kink interaction.

This simple calculation yields important insights, but at the same time, it is quite special and can be performed as such
only in the case of integrable models. Unfortunately, most
realistic applications do not yield integrable models, and are not amenable to methods based on exact solutions.
As a result, we must develop techniques
that work beyond the strict confines of integrable
settings.

\section{Then A Generalization: Nonintegrable Klein-Gordon Models}
\label{sec:nonintegrable}

Nonlinear Klein-Gordon PDEs are the natural generalization of the sG equation, taking the form:
 \begin{equation}
u_{tt}=u_{xx}-V'(u).
\label{nonlinKG}
  \end{equation}
 Many important examples have been studied, most famously the sG equation, stemming from the potential $V(u)=1-\cos(u)$ and  the $\phi^4$ model~\cite{ourp4}, for which $V(u) = \frac{1}{2}(1-u^2)^2$.
 The $\phi^4$-system will be at the epicenter of the next few sections, as attempts to understand its dynamics led to the mistake motivating this article. 
 The $\phi^4$ model is often considered as a prototypical
 system for phase transitions, ferroelectrics, high-energy 
 physics and other applications~\cite{gibbon,ourp4}.
 
 Importantly,  many of these models have exact
 solutions such as kinks; in the $\phi^4$ case, for example,
 they interpolate between the temporally stable fixed points
 of $u=+1$ and $-1$ (or vice versa) and have the explicit functional
 form:
 \begin{equation}
\uK(x,t)=\tanh{\left(\frac{x-x_0-v t}{\sqrt{1-v^2}}\right)},\,
\uKbar(x,t)=-\uK(x,t).
\label{nonlinKGa}
 \end{equation}

In~\cite{manton}, Manton constructed a method to characterize the interaction between kinks
 (and antikinks) that applies to Hamiltonian wave equations, both integrable and non-integrable.  We briefly review it here. Models such as Eq.~(\ref{nonlinKG}) conserve various quantities, i.e.\ leave them unchanged over time. These include
 the energy
 \begin{equation}
H=\int_{-\infty}^{\infty} \frac{1}{2} u_t^2 + \frac{1}{2} u_x^2 + V(u) dx
\label{eq5}
 \end{equation}
 and the momentum
 \begin{equation}
P=-\int_{-\infty}^{\infty} u_t u_x dx.
\label{eq6}
 \end{equation}
 If we  consider the momentum contained between two locations
 $x=a$ and $x=b$, rather than between $-\infty$ and $\infty$, we can
 directly derive, using Eq.~(\ref{nonlinKG}),
  \begin{equation}
\frac{dP}{dt} ={\left[-\frac{1}{2} u_t^2 -\frac{1}{2} u_x^2 + V(u)\right]}_a^b
\label{eq5aa}
  \end{equation}
  If we now assume an anti-kink at $x=0$ and a kink at $x=s$, then
  we can approximate the waveform of their superposition
  (more on this approximation later) as:
  \begin{equation}
u=u_1(x)+ u_2(x) + c, \quad u_1(x)=\uK(-x), \quad u_2(x)=\uK(x-s)
\label{eq5a}
  \end{equation}
  where the K subscript has been again used to 
  denote the kink nature of the waveforms.
  Then, under the assumption that $a \ll 0 \ll b \ll s$,
  \begin{equation}
  \frac{dP}{dt} \approx \left[-\frac{1}{2} u_{1x}^2 - u_{1x} u_{2x} +
V(u_1) + V'(u_1) u_2 \right]_a^b.
  \label{eq5b}
  \end{equation}
The contributions to the momentum change due to the
  interaction between the two waves stem from the 2nd and 4th terms and hence using the
  kink asymptotics $u_1 \approx -c + A \exp(-m x)$
  and $u_2 \approx -c + A \exp(m (x-s))$ (for sG $m=1$ and $A=4$), we obtain 
  \begin{equation}
  \frac{dP}{dt} = 
  2 A^2 m^2 \exp(-m s) \Rightarrow \Veff(s) = 
  -2 A^2 m \exp(-m s)
  \label{eq5c}
  \end{equation}
  leading to the equation of motion
  $\ddot{s}=-\frac{2}{M} \frac{dP}{dt}$ for the general Klein-Gordon case.
  In the sG case, this calculation produces the same result found in the previous section,
  while in the non-integrable $\phi^4$ case,  it gives $\ddot{x}=-16 \exp(-2 x)$.
  That is to say, well-separated kinks and antikinks still attract each other and would
  collide accordingly. Moreover, the above methodology of~\cite{manton}
  can be generalized broadly to nonlinear wave systems in which the forms of the waves  (or at least their asymptotics) are known.

  \section{The Curious Case of the $\phi^4$ Model: Collisions and Multi-Bounce Windows}

\begin{figure}[tbp]
\begin{center}
  \includegraphics[width=0.9\textwidth]{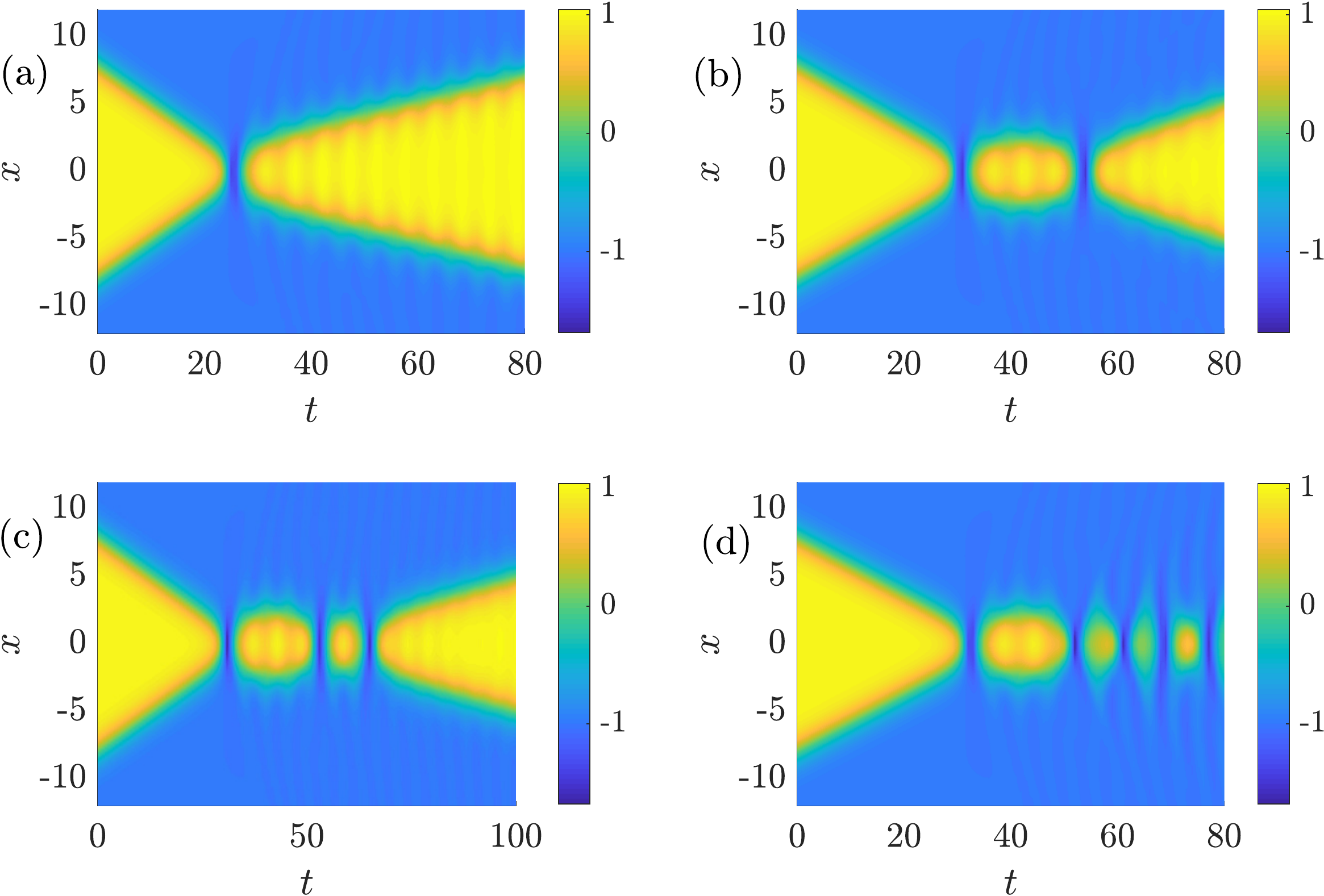}
  \caption{Simulations of $\phi^4$ kink-antikink solutions with varying initial velocity: (a) $v_0=0.28$, one-bounce solution, (b) $v_0=0.225$, a two bounce solution, (c) $v_0=0.2211$, a three-bounce solution, and (d) $v_0=0.21$, illustrating caption and bion formation.
}
\label{fig:phi4kinks}
\end{center}
\end{figure}

  Based on the above section, it is tempting to think that the collision dynamics of 
  non-integrable Klein-Gordon PDEs behave, up to small variations, just like their integrable
  counterparts. Compare the kink-antikink solutions for the sG system in Fig.~\ref{fig:SG2kinks}(a) with the kink-antikink solution to $\phi^4$ in Fig.~\ref{fig:phi4kinks}(a), both given with a relatively large initial velocity
 $v=0.28$. The two simulations seem roughly similar, but with a few nontrivial differences.  First note that since  $u=2\pi m$ is a stable equilibrium of sG for all $m$, kinks and antikinks  pass through each other following a collision, whereas in $\phi^4$, with only two stable equilibria, kinks and antikinks are reflected following a collision.
  One can also see that the sG kinks  collide  elastically: after the collision they pass through each other and continue with their original incoming speed. On the other hand,
  the $\phi^4$ kinks 
  bear the signature of the non-integrability of
  the model. In particular, their collisions are \emph{inelastic}
  and as a result the kink and antikink  lose kinetic energy in the collision, as is evident from the reduced slope in the space-time
  contour plot. A secondary effect is slightly harder to discern at first
  glance, but is all-important. Namely, after the collision, the kinks in the $\phi^4$ simulation seem to be slightly
  ``wobbly.''  This, we will see momentarily, is due to an internal vibration of the coherent structures. 
  
However, simulations of the $\phi^4$ model with lower initial speeds result in \emph{dramatic departures} from the elastic
(irrespectively of the kink speed) collisions seen in the integrable sG model.  Fig.~\ref{fig:phi4kinks}(b) and (c) show the results of collisions with $v=0.225$ and $v=0.2211$ respectively. These demonstrate, respectively, what are now known as a two-bounce solution, in which the kink and antikink collide, begin to separate and then re-collide before ``escaping'' each other's attraction, and
a three-bounce solution, which features one more round of separation and re-collision than the two-bounce solution.
In fact, if the speeds of the waves are sufficiently low, then they are
never able to escape each other's attraction and persist in a
long-lived trapped waveform which is termed a ``bion'', as is shown in Fig.~\ref{fig:phi4kinks}(d). The keen-eyed will observe the presence of radiation propagating ahead of the escaping kink-antikink pairs in all these images. Obviously,
such features are unprecedented, and indeed impossible, in
integrable systems. 

In the late 1970's and into the
1980's, researchers became intrigued by the 
deviations between the dynamics in  $\phi^4$ and related non-integrable models relative to those described by the IST for integrable systems and began to appreciate their relevance
in mathematics and physics. Early numerical
studies by numerous researchers---Kudryavtsev~\cite{kud}, Aubry~\cite{aub},
Getmanov~\cite{get}, among others---reported the possibility of trapping, but 
arguably  Ablowitz, Kruskal and Ladik~\cite{AKL}
 first reported the possibility of multi-bounce solutions. Numerical simulations back then were slow, expensive, and difficult to visualize. Each of these studies reported on a small handful of simulations. 
 
\begin{figure}[tbp]
\begin{center}
  \includegraphics[width=0.45\columnwidth]{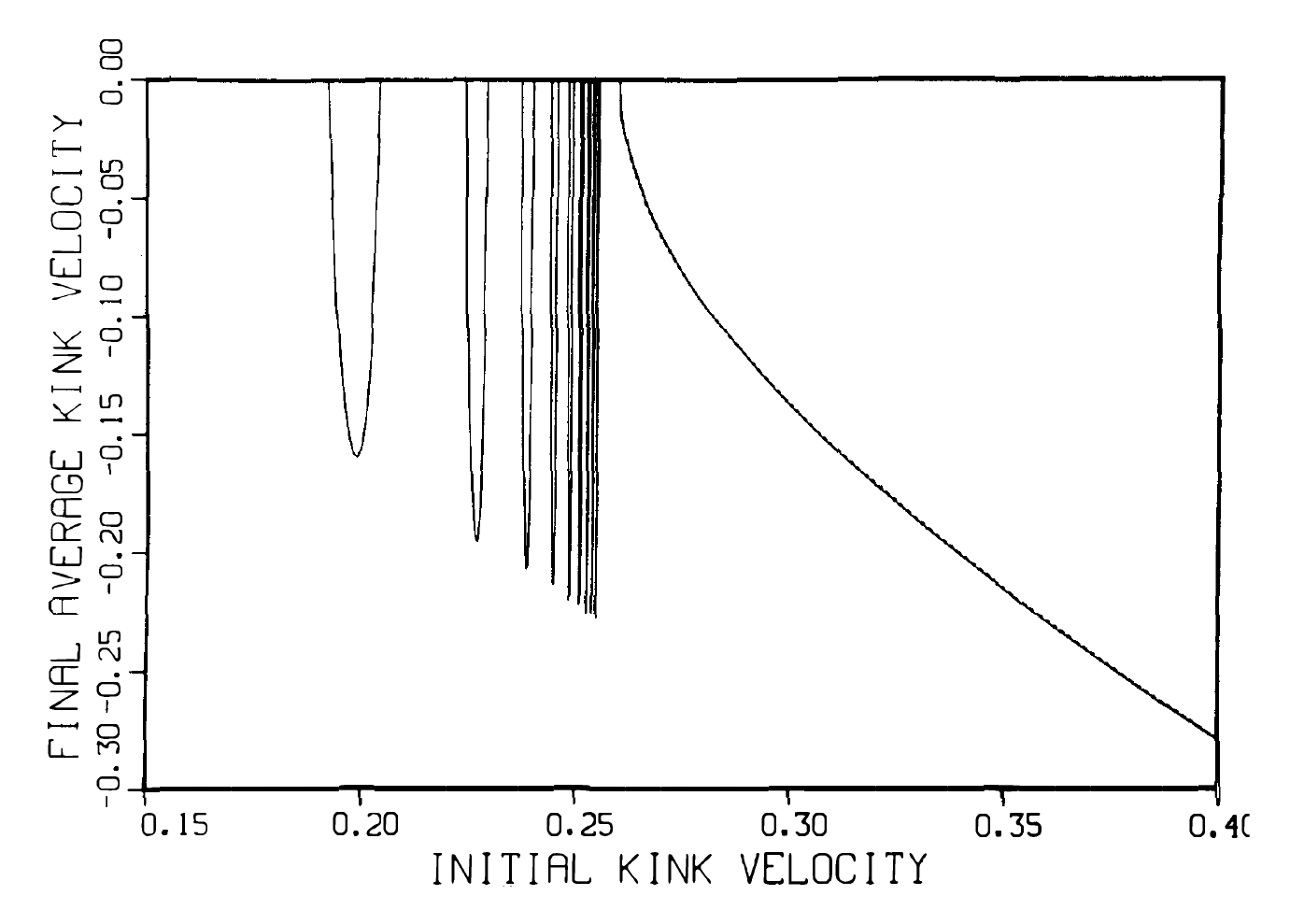}
  \includegraphics[width=0.45\columnwidth]{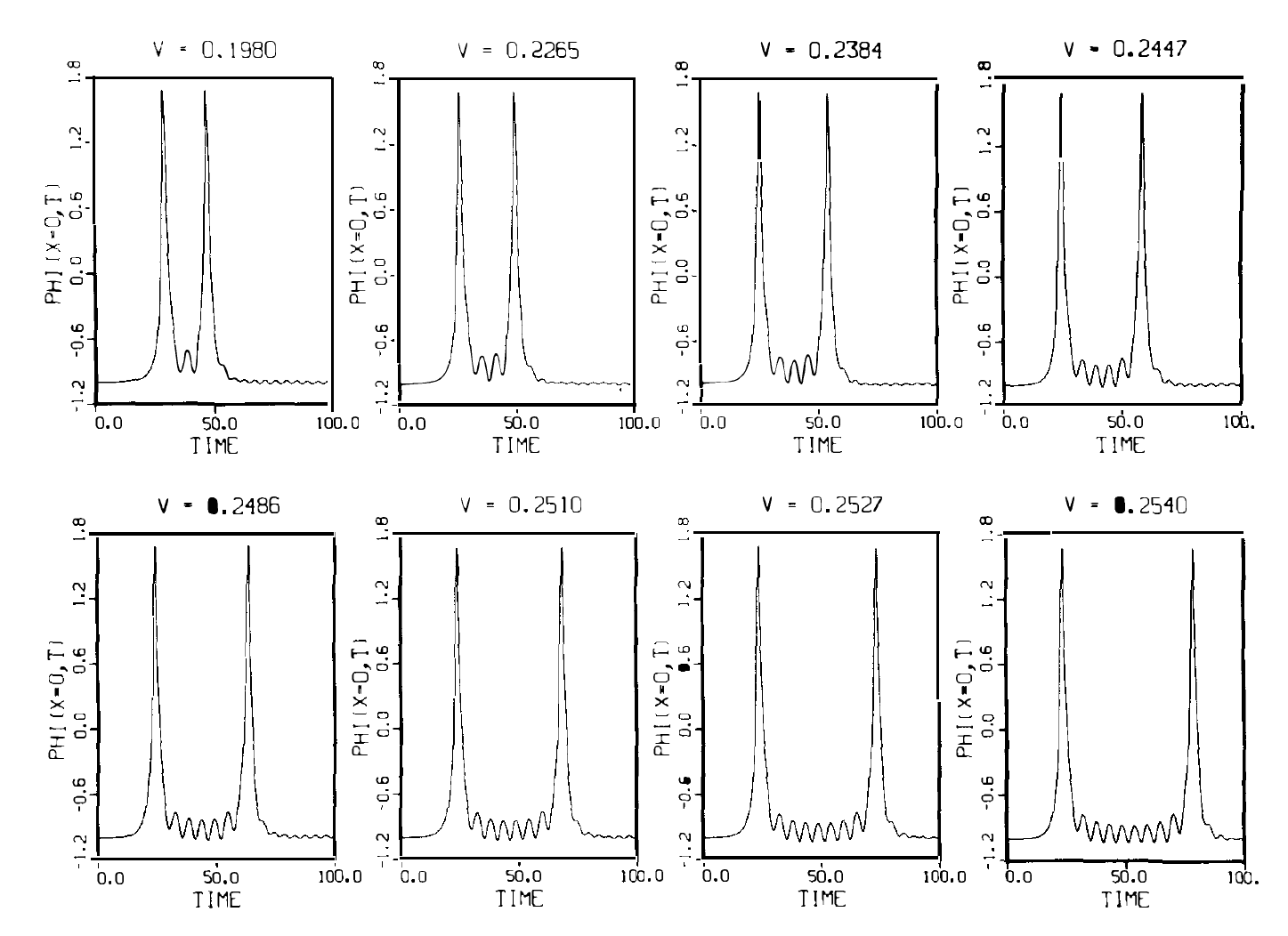}
  \caption{Typical examples from the work of~\cite{csw} as adapted in~\cite{ourp4} (used with
permission from the latter). 
The left panel shows the area of two-bounce windows
of outgoing vs.\ incoming velocity, before the single-bounce
occurring after the critical velocity of $\vc \approx 0.2598$.
The right panel shows how progressively more vibrations of the internal
mode occur when probing the field at the origin $u(0,t)$ as a function
of time for select speeds in the different two-bounce windows.
}
\label{fig2}
\end{center}
\end{figure}

The first to realize the complexity of these collisions were D. Campbell and his group at Los Alamos, who performed the first reasonably thorough numerical study, systematically tabulating the value of the outgoing collision velocity
 as a function of incoming collision velocity~\cite{csw}. This is reproduced in the left panel of Fig~\ref{fig2}. They found that at initial velocities above a critical value $\vc$, the kink and antikink escape after a single inelastic collision. Below this value, they found that the two-bounce solutions occur in a sequence of ``windows'' of finite width. At the center of each window is a ``resonant velocity'' at which the output speed nearly equals the input speed. They did not report any three-bounce solutions. The right panel shows the value of $u(0,t)$ for simulations of collisions  at each of the resonant velocities. The two collisions occur at the large maxima, and each shows one more ``bump'' between the two collisions than the one before it. These simulations used so much computational resources that Campbell's supervisor at Los Alamos, A. Scott, had to check in with him to make sure it was necessary~\cite{Scott2005}.
D. Campbell
has written a very accessible overview of these first efforts in
the first Chapter of~\cite{ourp4}.

The key insight of~\cite{csw} is that  two-bounce windows are made possible due to the presence of an internal mode
in the spectrum of the $\phi^4$ kink. Indeed, the spectrum of 
the sG equation linearized around  the sG kink  consists solely of a pair of
zero eigenvalues (corresponding to the kink's translational invariance)
and the continuous spectrum (associated with the background states
between which the kink interpolates). However, in the $\phi^4$ case,
the corresponding linearization yields a P{\"o}schl-Teller (i.e.\ $\sech^2$) potential that has an
additional bound state with frequency $\ws=\sqrt{3}$
and eigenfunction (for a static kink centered at $x_0$):
\begin{equation}
 u = \sqrt{\frac32} \tanh(x-x_0) {\rm sech}(x-x_0).
 \label{internalmode}
\end{equation}
The factor $\sqrt{\frac32}$ is chosen to normalize the solution for convenience in a later calculation.
This internal mode provides the system with the ability  to store
potential energy in a neighborhood of the kink. Hence, the main idea  is:
\begin{itemize}
\item The incoming kinks have (kinetic) \emph{energy} $K=\frac{M}{2}\vin^2$ where the mass of the solitary wave is given by~\eqref{kinkMass}. The kink and antikink induce a potential corresponding to their mutual attraction, as discussed in previous sections.
\item Some of this energy, upon collision, is converted to potential
  (internal) energy of this vibrational mode.
\item Between the emission of small amplitude radiation (after all, the origin of inelasticity  of the collision in a non-integrable model; one of the miracles of integrable systems is that collisions generate no such radiation) and the depositing
  of energy partly in this internal mode, the kinks are left with
  low enough kinetic energy that they are \emph{unable} to escape
  each other's attraction.
\item As a result they only separate to a maximal distance, and then
  return and re-collide.
\item At the second collision, the exchange of energy may go in either direction and depends both on the amplitude of the oscillatory mode \emph{and its phase}.
\item If the phase of the oscillatory is such that it returns sufficient energy to the propagating mode for the kink and antikink to escape their mutual attraction, the process is called \emph{resonant}.
\item If the kink and antikink do not escape on the second bounce, they may escape due to resonance on a later one.
  \item Because each collision generates radiation, in addition to exchange between the potential and kinetic modes, with each collision, the probability of eventual escape decreases. If enough energy is converted to radiation, a bion state will form.
\end{itemize}
Subsequent numerical studies by Anninos et al.~\cite{anni91} and by Goodman and Haberman~\cite{good07} showed the existence of additional narrower windows corresponding to three or more bounce resonances. A  recomputation of Fig.~\ref{fig2}, using more modern numerical methods, is given in Fig.~\ref{fig3}, from Goodman's contribution to Ref.~\cite{ourp4}, showing all computed $n$-bounce windows up to $n=5$.

The resonance condition posited by~\cite{csw} was that the internal (shape)
mode frequency $\ws$ and the interval $T_2$ between subsequent collisions needs to satisfy $\ws T_2= 2 n \pi + \delta$,
where $\delta$ is some offset phase to be found by a fit. Such windows were found numerically for $3\le n \le 10$. Using particle
mechanics, the authors of~\cite{csw} were able to connect $T_2$ with the difference between the kinetic
energy and the critical kinetic energy for kink-antikink separation
according to $T_2 \propto {\left(\vc^2-v_n^2\right)}^{-\frac{1}{2}}$
This phenomenological approach finally led them to an
empirical formula for the $n$th two-bounce window; 
see also the right panel of Fig.~\ref{fig2}.
The relevant formula reads:
\begin{equation}
 v_n^2 = \vc^2- \frac{\alpha}{{(2 \pi n + \delta)}^2},
 \label{eq:vn}
\end{equation}
where $\alpha$ is an additional parameter determined by fitting. 
This was found to agree remarkably well with the location of the
first 10 two-bounce windows! It provides a phenomenological explanation why there are no windows with $n=1$ or $n=2$ as these would return a negative value of $v_n^2$ given the empirical values of $\alpha$ and $\delta$. Importantly, subsequent work of
Campbell and collaborators showed that
other non-integrable models such as the modified
sine-Gordon equation~\cite{peyrard},
the double
sine-Gordon equation~\cite{sodano},
and others featured a similar
phenomenology.

\section{A Simplifying Approximation and A Crucial Mistake}

The approach of~\cite{csw} is quite useful in unveiling the
principal mechanism of the two-bounce resonance, yet it is rather phenomenological
in that it depends on the numerically-determined parameters $\alpha$ and $\delta$ in Eq.~\eqref{eq:vn}.
This indicates that it is based on an incomplete mathematical understanding of the problem. A more quantitative theoretical framework is needed to avoid such dependence on fitting. Fortunately, even before the publication of~\cite{csw}, Sugiyama~\cite{sugi79} derived a simplified system of model equations that sought to provide such a framework. This framework has two advantages over Manton's computation, described in Sec.~\ref{sec:nonintegrable}: first, it makes no assumption of large separation, an assumption violated by the colliding pair, and, second, it takes into account the transfer of energy to the secondary mode~\eqref{internalmode}.  Sugiyama's approach has been at the heart of numerous inquiries into the two-bounce phenomenon over several decades~\cite{sugi79,kudr87,anni91,good07,good15}.

Sugiyama used the so-called variational method to derive a finite dimensional system of ordinary differential equations modeling the kink-antikink interaction. The method is often attributed to Bondeson et al.~\cite{Bondeson:1979bh}, but this paper was submitted a mere three days before Sugiyama's so the origin of the idea remains (for us) cloudy. The method applies to PDE systems derived from a variational principle, meaning that the PDE arises as the Euler-Lagrange equations minimizing a certain action
\begin{equation} \label{Lagrangian}
A(u) = \int L(u;t)\, dt = \iint \cL(u(x,t))\, dx\, dt.
\end{equation}
For example the nonlinear Klein-Gordon models~\eqref{nonlinKG} have a Lagrangian of the form
  \begin{equation}
L=\int_{-\infty}^{\infty} \frac{1}{2} u_t^2 - \frac{1}{2} u_x^2 - V(u)\, dx.
\label{KGlagrangian}
 \end{equation}
The method is simple: assume that the solution depends on a finite number of time-dependent parameters $a_1(t),\ldots,a_n(t)$
\begin{equation}\label{ansatz}
u(x,t)=U(x;a_1(t),\ldots,a_n(t)).
\end{equation}
Substituting this ansatz into Eq.~\eqref{Lagrangian}, and integrating with respect to $x$ yields a finite-dimensional Lagrangian whose Euler-Lagrange equations yield an ODE system for the evolution of the parameters $a_j(t)$. The method yields equations for the minimizer of the action among all functions in the multi-parameter family~\eqref{ansatz}. 

The fidelity of the resulting dynamics can be no better than the quality of the guess, which in turn depends on the modeler's understanding of the system. For example, one could try an ansatz involving only the translational mode. In that case,
the variational \emph{trial solution} will have the form:
\begin{equation}
  u=\uK(x-X(t)) + \uKbar(x+X(t)) -1.
  \label{ansatzKinkOnly}
\end{equation}
When this is substituted in the field Lagrangian of the $\phi^4$ model,  the effective Lagrangian and equation of motion take the form:
\begin{equation}
L(X,\dot{X})=a_1(X) \dot{X}^2 - a_2(X) \Rightarrow
\ddot{X}=-\frac{1}{2a_1} \left[a_1'(X) \dot{X}^2 + a_2'(X) \right].
\label{eq5h}
\end{equation}
Here the quantities $a_{1,2}(X)$ denote suitable overlap
integrals---e.g.\ it is straightforward to obtain that
$a_1(X)=(1/2) \int_{-\infty}^{\infty} {\left(\uK'-\uKbar'\right)}^2 dx$, where 
 the prime denotes the derivative with respect
to the function's argument. Given that the single dynamical
equation in Eq.~(\ref{eq5h}) conserves energy, the kinks in this ``highly constrained''
reduction of the infinite-dimensional (PDE) dynamics can do nothing but collide elastically. Using this ansatz eliminates the need to assume the kink and antikink are well separated but provides too few degrees of freedom to allow the observed phenomena in the case of the $\phi^4$ model.

We  can achieve greater fidelity by including an internal mode in the ansatz. 
This permits an energy exchange between the total energy (kinetic plus potential) and the additional potential energy stored in the internal vibration around the kink. In this case, the trial solution reads
\begin{equation}
  u=\uK(x-X(t)) + \uKbar(x+X(t)) -1 +  \left[A(t) \chi(x+X) + B(t) \chi(x-X)\right]
  \label{ansatzKinkPlusInternal}
\end{equation}

leading to the far more complicated Lagrangian of the form:
\begin{equation}
L=a_1 \dot{X}^2 -a_2(X) + a_{31} (\dot{A}^2 + \dot{B}^2) -a_4 (A^2 + B^2)
+ a_{32} \dot{A} \dot{B} - a_{42} AB + a_5 (A-B) + \dots
\label{ABXlagrangian}
\end{equation}
This Lagrangian only contains what are thought to be ``essential terms''
of interaction, assuming that the amplitude of the internal modes
is small enough that higher powers of $A$ and $B$ can be neglected or are
secondary to the (up to quadratic) powers considered herein. Indeed,
the work of Weigel and his group~\cite{wei1,wei2,wei3} reports
on these higher powers too for completeness.

The crucial error of the work of~\cite{sugi79} already occurred
at this level. The expression for $a_5$ should read:
\begin{equation}
  a_5(X)=-3 \pi \sqrt{\frac{3}{2}}
  \left[2-2 \tanh^3(X)-3 {\rm sech}^2(X)+{\rm sech}^4(X)\right].
  \label{eq5k}
\end{equation}
However, Sugiyama must have inadvertently changed the $\tanh^3$ to $\tanh^2$, which allows the bracketed term to be simplified to $-\sech^2{(X)}\tanh^2{(X)}$. This, sadly,
was the expression effectively used throughout the literature. In the work discussed below, the Lagrangian is simplified by assuming $A=-B$, equivalent to assuming that the solution lies on the invariant manifold of solutions satisfying even symmetry $u(x,t)=u(-x,t)$. That is, one can easily check that solutions to~\eqref{ABXlagrangian} with $A(0)=-B(0)$ and $\dot{A}(0)=-\dot{B}(0)$ satisfy $A(t)=-B(t)$ for all $t>0$.

\begin{figure}[htbp]
\begin{center}
\includegraphics[width=.9\textwidth]{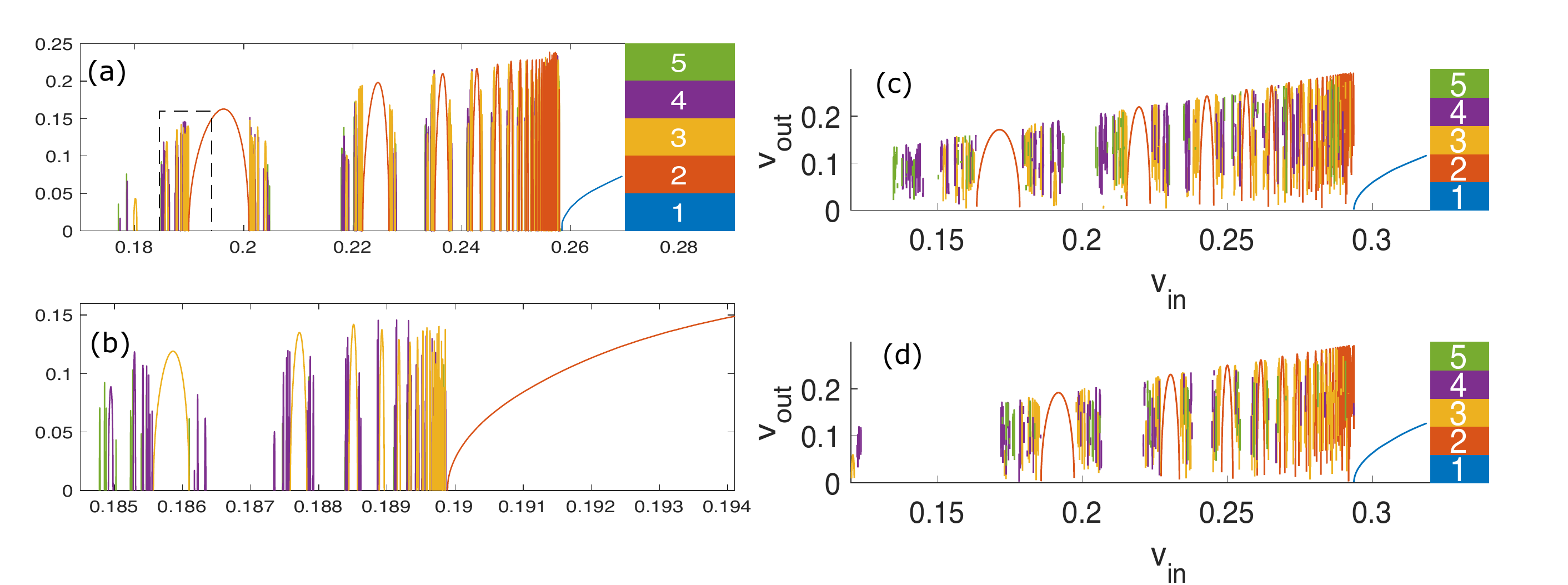}
  \caption{Modern renderings of the results of Fig.~\ref{fig2} showing (a) results of PDE simulations, (b) zoomed-in view of the dashed box above it, demonstrating fractal-like structure, (c) equivalent image from the qualitative ODE model, (d) equivalent image generated by discrete map approximation to ODE model; adapted with permission
  from~\cite{ourp4}.}
\label{fig3}
\end{center}
\end{figure}

This mistake propagated from~\cite{sugi79} to all the crucial works
that followed  considering the analysis of this model including,
e.g.,~\cite{kudr87,anni91,good07,good15}. Anninos et al.\ showed numerically that the two-degree of freedom ODE system possesses a fractal like structure, with three-bounce windows accuumlating near the edges of the two-bounce windows, four-bounce windows at the edges of the three-bounce windows, etc; Fig~\ref{fig3}(b).

Goodman and Haberman followed this up with a comprehensive analysis of Sugiyama's ODE system~\cite{goodman2005}. Using phase plane analysis, Melnikov integrals, and matched asymptotics expansions, they derived analytical expressions for the critical velocity $\vc$ as well as the parameters $\alpha$ and $\delta$ given in equation~\eqref{eq:vn} (without recourse to parameter fitting), as well as similar formulas for the locations of the three-bounce windows.  Unfortunately, this analysis was performed on the ODE with the problematic $a_5$ term. Subsequently, they extended this analysis to reduce the ODE system to a discrete-time iterated map, for which they  derived a detailed bifurcation diagram~\cite{goodman2008}. For purposes of exposition, they worked in this paper with an idealized ODE model that retained the essential dynamical features of Sugiyama's ODE system, while having a somewhat simpler structure. Such a map reproduces in great detail much of the fractal structure seen in the ODE model; see Fig~\ref{fig3}(d). 

In addition to noticing the problem with the $a_5$ term and correcting it, Weigel's group pointed out that the qualitative and semi-quantitative match disappears when the corrected form of $a_5$ is used~\cite{wei1,wei2,wei3}. They further showed, to make matters worse, that the disagreement between ODE and PDE results only increases when higher order terms neglected in~\eqref{ABXlagrangian} are included.
One final problem makes make matters even gloomier, the so-called null-vector singularity. The internal mode~\eqref{internalmode} moves in a potential defined by the kink. However at $X=0$ the term~\eqref{ansatzKinkOnly} representing the kink and antikink in the more general ansatz~\eqref{ansatzKinkPlusInternal} vanishes, rendering the concept of an internal mode meaningless. This fact arises in the context of the evolution equations by making the mass matrix associated with
the dynamical evolution singular when $X=0$. Such a singularity
is absent in the original PDE and is a mere side effect of the ODE reduction method; Caputo et al.\ recognized this problem in the early 1990's and showed how to remove the singularity via a nonlinear change of variables~\cite{caputo1991,caputo1994}. 

To avoid such issues, Weigel's group~\cite{wei1,wei2,wei3} proposed the following modified ansatz in the variational trial function:
\begin{equation}
  u=\uK(x-X) + \uKbar (x+X) -\tanh(q X) +  \left[A \chi(x+X) + B \chi(x-X)\right].
  \label{ansatzWeigel}
\end{equation}
This introduces a repulsive potential in the vicinity
of $X \rightarrow 0$, which, in turn, precludes the kink and
antikink from hitting the singularity at $X=0$. Of course, this
now adds an artificial potential and  $q$ becomes a tunable
parameter which must be chosen optimally in order to optimize the fidelity
of the reduced model to the original PDE. This can be done
in a variety of ways, including selecting $q$ to capture the
right PDE outgoing velocity, or in order to capture the right
number of bounces, or to possibly minimize the distance from
the PDE kink-antikink center trajectory, or satisfy some
other suitable cost function criterion. Whichever way is selected though
cannot bypass the fact that this is a phenomenological
and seemingly artificial inclusion that cannot be made
systematic.

Remarkably, this suggests that despite a tremendous effort
and $40$ years of significant developments, we are still
missing a quantitative understanding of the relevant phenomenology
of what is arguably the simplest non-integrable collision dynamical
model (the $\phi^4$ model), in the context of the simplest
type of coherent structures (heteroclinic, real-valued kinks
with only one internal mode). We highlight this because the examples
in the following section are more elaborate
and have more complex and tunable features. This appears to be a disaster, but we choose, instead, to  view it as an opportunity. 

Clearly,
we have amassed a tremendous amount of experience about the
relevant phenomenology. The fractal-like structure of multi-bounce windows
is well-established, and we know that it arises due to the (nearly) reversible transfer
of energy between translational and vibrational modes.
This is known to be due to the internal mode.
Yet, we still lack a systematically-derived model
that quantitatively characterizes this structure.

Perhaps a relevant suggestion in this direction
is a beyond-two-mode ansatz that properly incorporates not only
the translational and internal modes, but also the potential
of the kink to radiate energy through modes of the continuous
spectrum. This irreversible transfer to background modes is
one of the significant features lacking in all of the earlier
considerations. A formal analysis including the leading-order effects
was used to derive a nonlinear damping term in a similar variational model 
describing the interaction of a sine-Gordon with a delta-function potential in~\cite{ghw2002}. The effects of adding such a term to the variational model equations for the $\phi^4$ model are discussed in~\cite{good15}.

A deeper question concerns the variational method itself. Despite its popularity in the mathematical physics literature going back at least 40 years, we know of only a single instance of a mathematical theorem proving that it works for any particular problem. That is, we know of only one paper in which an approximation derived using the variational method has been shown to remain close to the full solution of a PDE model over a time period of interest. In contrast to methods such as averaging or multiple scales, the variational approximation is not a perturbation method; it does not depend on a small parameter, so that applying the method does not yield an explicit remainder term which can afterward be shown small with respect to that parameter. By contrast, it is standard procedure to first apply a formal perturbation method such as averaging and to justify afterward it analytically. Chong et al.\  show that variational approximations to time-periodic orbits in a discrete NLS become increasingly accurate as a small parameter in that problem is reduced~\cite{chong2012}. This result both depends on a small parameter and makes use of the time-periodicity of the solutions and thus does not generalize to the problem being discussed which is a general initial value problem with no small parameters. Chong's work follows an earlier non-rigorous result due to Kaup and Vogel~\cite{Kaup}, which bounds the error due to the method but only to the special case of steady solutions, in which case the variational method yields algebraic equations.
Therefore, we make a plea to our analyst colleagues: explain why and when the variational method works, or help us replace it with a justifiable method!

 \section{Recent Developments: Beyond the $\phi^4$ Model}

 In addition to recent developments  in the $\phi^4$ realm,
 the last decade has witnessed the discovery of intriguing features in higher-order
 models. One of the
 most well-established such models is the $\phi^6$
 equation, which arises in the work of
 Dorey \emph{et al.}~\cite{dorey}. The potential in its Lagrangian~\eqref{KGlagrangian} takes the form:
 \begin{equation}
V(u) = - \frac{1}{2} u^2 (u^2-1)^2.
\label{eqn6}
 \end{equation}
 The system possesses equi-energetic minima $u= -1, 0, 1$, as 
 well as  kink solutions that connect the various pairs of 
 minima, such as, 
 \begin{equation}
u=\pm \sqrt{\frac{1+\tanh(x)}{2}},
\label{eqn7}
 \end{equation}
 and which can be Lorentz boosted to construct moving kinks.
 Remarkably, kink-antikink collisions in this model display resonance windows, despite the fact that the linearization about these kinks contains no internal modes: their only discrete eigenmodes correspond to the translation invariance of the underlying equations. These authors make the groundbreaking qualitative observation that, instead of considering the spectrum of an isolated kink, one must consider the spectrum of the linearization about the  \emph{dynamic
configuration} of a kink--antikink pair moving towards each other.
 They find that this linearized system possesses a (transient)
 internal 
 mode  which they claim is responsible for
 the resonance phenomena. This is a bold proposal that requires
 some firm mathematical ``backing''. It poses the more
 general question of the potential meaning of spectral features
 around a time-dependent state for which there is no frame where it
 can be considered stationary. In dissipative systems where the
 spectral gap may allow only for a few modes to be dynamically
 relevant, this may be easier to justify, but understanding the
 meaning of such a spectral analysis in Hamiltonian, conservative
 problems is a \emph{wide open} field for future consideration.

\begin{figure}[tbp]
\begin{center}
  \includegraphics[width=\textwidth]{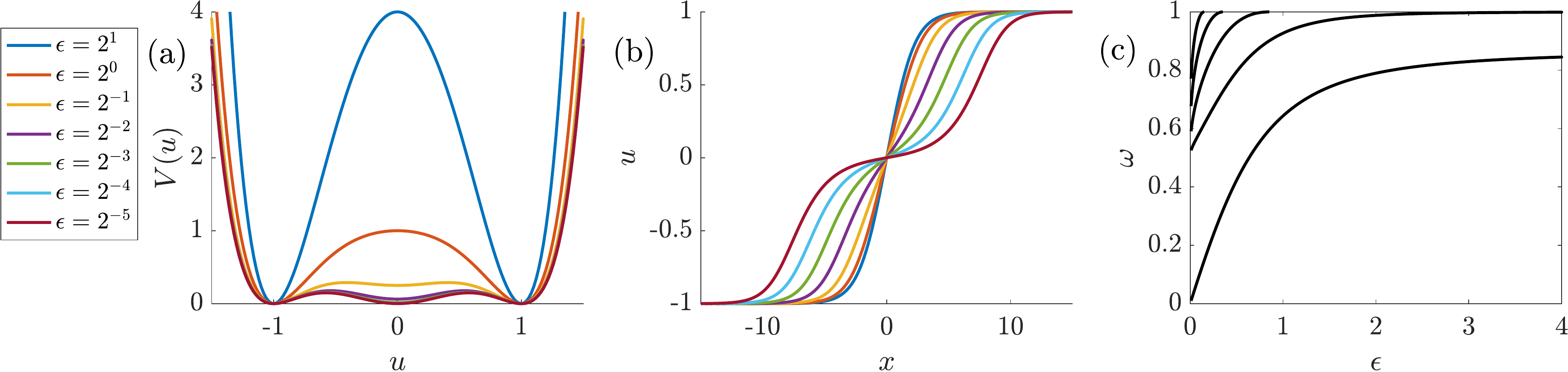}
  \caption{Typical results from the $\phi^6$ model of~\cite{demirkaya}. Legend shared by subplots (a-b).
(a) The potential $V(u)$
for different values of $\epsilon$---for larger $\epsilon$ it behaves like $\phi^4$, while as $\epsilon \searrow 0$ it approaches a triple well with three equal minima.  (b) The corresponding kinks, approaching the $\tanh$ kink of $\phi^4$ for large $\epsilon$ and approaching a two-step structure
as $\epsilon$ decreases. (c) The discrete spectrum as a function of $\epsilon$, showing two eigenvalues for large values of $\epsilon$ and additional eigenvalues bifurcating from the band edge at $\omega=1$ as $\epsilon$ is decreased.
}
\label{fig4}
\end{center}
\end{figure}

 Remarkably, this is not the only $\phi^6$ model that has yielded new results. 
 Demirkaya et al.~\cite{demirkaya} have recently considered a different $\phi^6$ model 
 proposed by Christ and Lee in 1975 (!) as a ``bag model'' where the domain walls
(kinks) play the role of quarks within hadrons in high-energy physics~\cite{Christ:1975db}.
They consider the potential:
\begin{equation}
V(u)=\frac{1}{8 (1+\epsilon^2)} (u^2 + \epsilon^2) (1- u^2)^2.
\label{eqn2}
\end{equation}
Remarkably and despite its functional complexity, the model features
an exact  (static) kink solution:
  \begin{equation}
u=\frac{{\rm sinh}\left(\frac{x}{2}\right)}{\sqrt{1+\epsilon^{-2}
+ \sinh^2\left(\frac{x}{2}\right)}}.
\label{eqn3}
\end{equation}
  Perhaps even more importantly, the linearization around this kink
  features a controllable number of internal vibrational modes, with a single mode in the
  $\epsilon \rightarrow \infty$ limit, where this model reduces to
  $\phi^4$, and an increasing number of modes, which is unbounded
  as $\epsilon \rightarrow 0$.  The relevant potential,
  solitary waves and linearization features
  are shown in Fig.~\ref{fig4}. The work of~\cite{demirkaya} showed that this  $\phi^6$ model possesses fewer multi-bounce windows as $\epsilon$ was
  decreased. 
  This makes intuitive sense: a larger number of internal modes allows the energy to split into more pieces, making it less likely that on subsequent collisions, they will simultaneously return their energy to the propagating mode.

More recent studies have considered models involving higher order potentials, namely the $\phi^8$, $\phi^{10}$ and $\phi^{12}$ defined by 
\[
V(u)=u^{2 n} (1-u^2)^2,
\]
for $n=2$, $3$, and $4$ respectively.
This is not just ``more of the same''; such models hold substantial mathematical
promise in their own right and are physically motivated by structural phase transitions in materials
science~\cite{ivan}. Unlike previously-considered systems, they support \emph{power law tails}. Since the kink  stationary
state satisfies: $\frac{du}{dx}=\sqrt{2V(u)}$, the following decay property calculation holds for the tails~\cite{asli2}. Suppose that as $x\to -\infty$, $u\to \bar{u}_i$ at algebraic rate $k_i$, and (respectively as $x\to\infty$, $u-u_{i+1}~\sim x^{-k_{i+1}}$), then, letting
\begin{equation}\label{eq:potzeros}
V(u)=\left(u-\bar{u}_i\right)^{k_i}\left(u-\bar{u}_{i+1}\right)^{k_{i+1}}
V_1(u),
\end{equation}
\begin{equation}\label{eq:int_bps}
\int dx=\int\frac{du}{\left(u-\bar{u}_i\right)^{k_i/2}\left(\bar{u}_{i+1}-u\right)^{k_{i+1}/2}\sqrt{2V_1(u)}}.
\end{equation}
\begin{equation}\label{eq:int_x}
\int dx\approx\frac{1}{\left(\bar{u}_{i+1}-\bar{u}_i\right)^{k_{i+1}/2}\sqrt{2V_1(\bar{u}_i)}}\int\frac{du}{\left(u-\bar{u}_i\right)^{k_i/2}}.
\end{equation}
Hence, e.g., for the $\phi^8$ model,
a kink connecting $-1$ and $0$ has asymmetric tails of the form:
\begin{align}\label{eq:kink1_asymp_plus}
u_{(-1,0)}(x)& \approx-\frac{1}{\sqrt{2}\: x} \,,\qquad x\to +\infty; \\
\label{eq:kink1_asymp_minus}
u_{(-1,0)}(x)&\approx-1+\frac{2}{e^2}\: e^{2\sqrt{2}\: x},\qquad x\to -\infty.
\end{align}
and similarly for the kink between $0$ and $1$.

\begin{figure}[tbp]
\begin{center}
  \includegraphics[width=0.35\columnwidth]{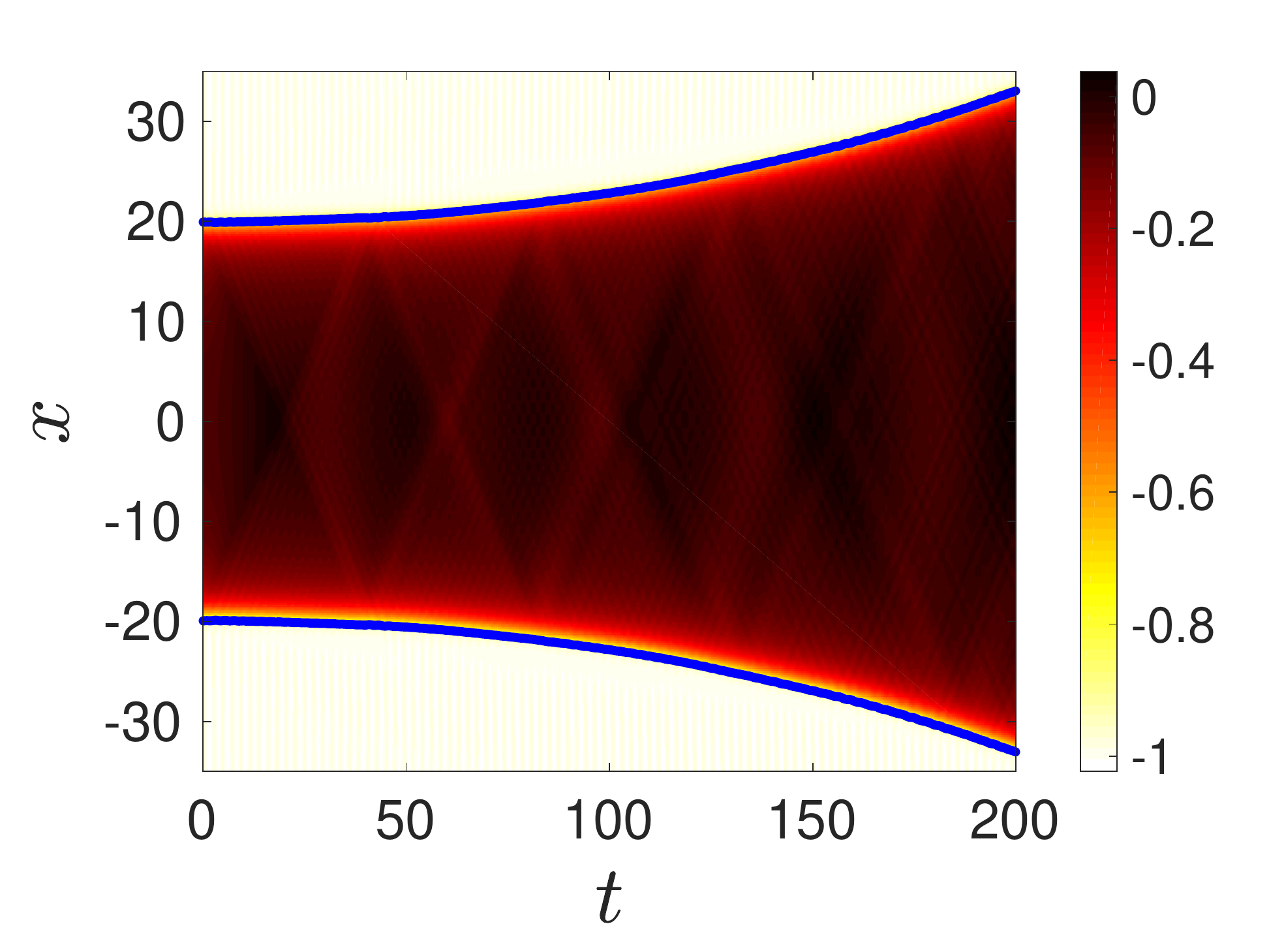}
  \includegraphics[width=0.35\columnwidth]{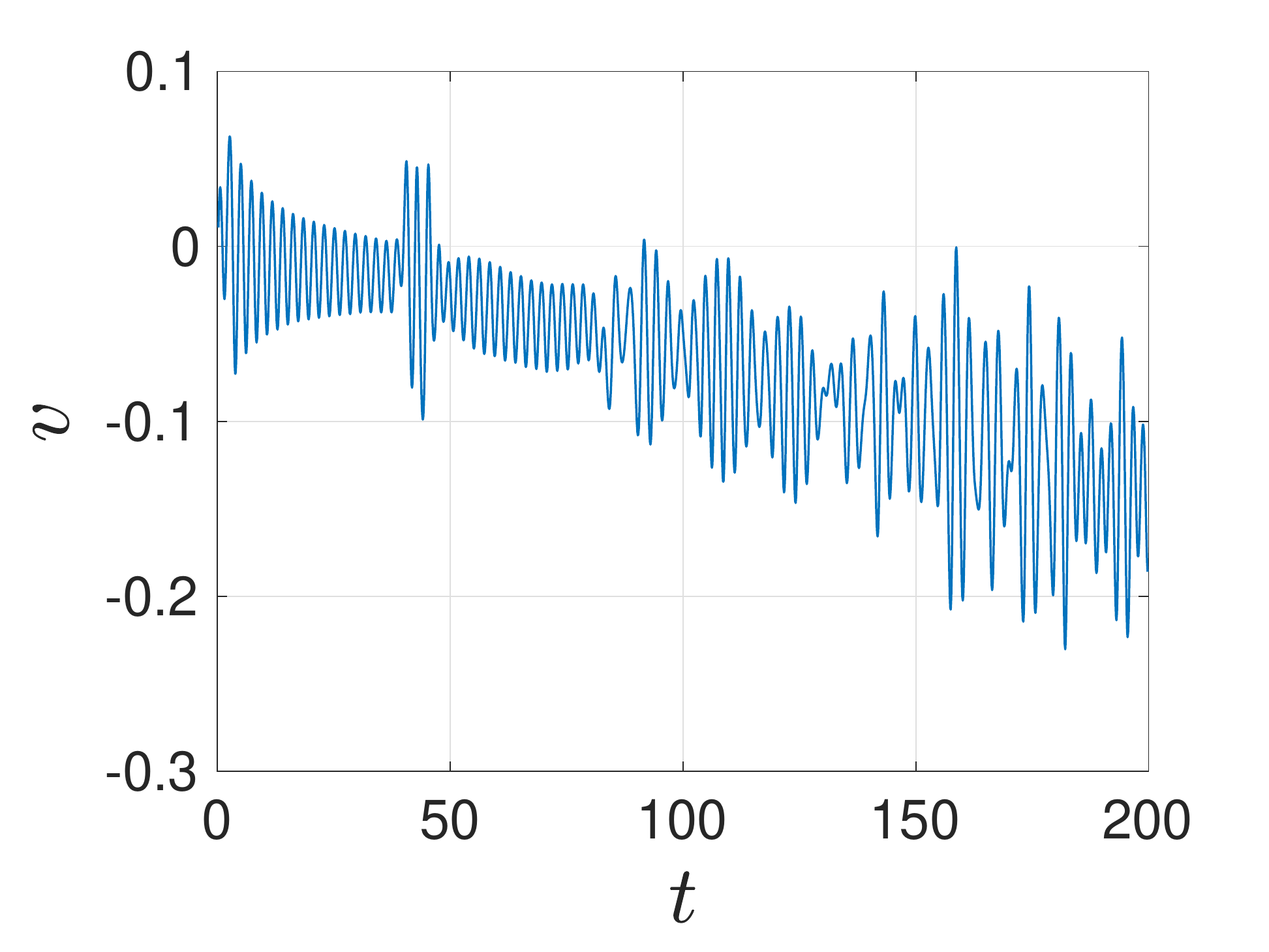}
  \includegraphics[width=0.35\columnwidth]{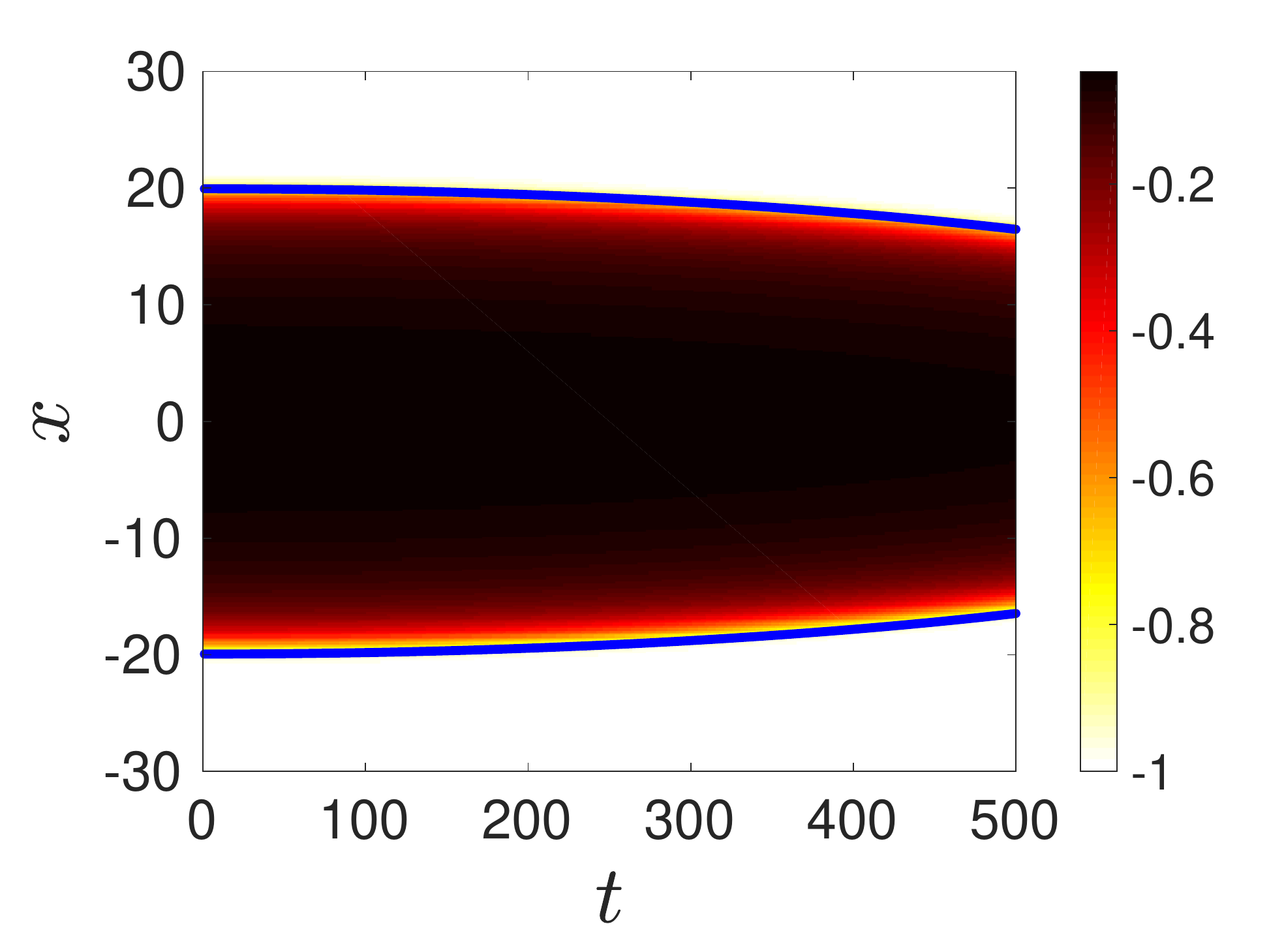}
  \includegraphics[width=0.35\columnwidth]{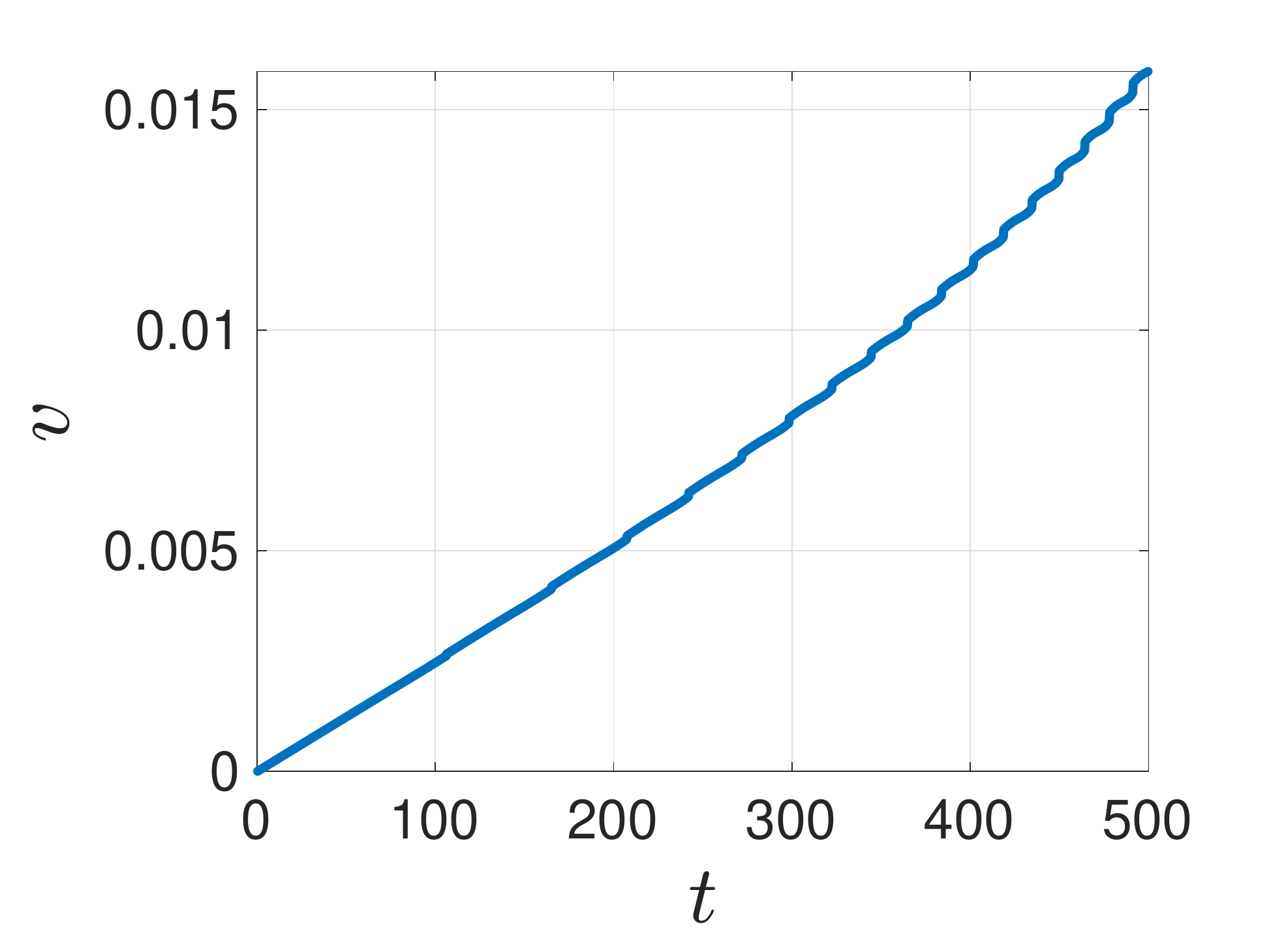}
  \caption{The unexpected impact of fat tails in the $\phi^8$ model:
a sum ansatz incorrectly seems to lead to kink-antikink repulsion
(and to a highly noisy velocity pattern) in the top panels.
The bottom panel showcases the result of the ``vacuuming'' procedure
(explained in the text) applied to the same ansatz, leading to a
much less noisy (as inferred by the velocity evolution of the
right panel) and indeed, as expected, attractive kink-antikink
interaction. Adapted from~\cite{asli2}.}
\label{fig5}
\end{center}
\end{figure}

This power law decay of the kink has numerous implications. A rather
unexpected one arose in the attempt to explore collisions:
a numerical initial condition consisting of a superposed a kink and an antikink with these
fat tails, using either a sum or a product superposition, simply does not work.
Fig.~\ref{fig5} shows the result of such a naive attempt: the numerics seems to suggest repulsion between the kink and antikink. Yet, that conclusion is in contrast with a theoretical calculation given below. It stems from the fact that the long-range tail of one of the structures dramatically
modifies the field vicinity of the other, skewing
the ``true'' interaction between the coherent structures  in an undesirable way.
The problem is that, because of the tails' slow decay, the superposition of the two widely-spaced and stationary solitary waves does not have small acceleration $\phi_{tt}$ at all points $x$. If instead, a minimization procedure is used to create an initial condition resembling such a superposition, but with $\phi_{tt}$ small everywhere, then the ensuing simulations do indeed produce the expected behavior.
The bottom panel of Fig.~\ref{fig5} shows the dramatic impact
of this procedure in unveiling the true, attractive nature of the
interaction between a kink and an antikink in the $\phi^8$ model,
contrasting the
apparent repulsion shown in the top panel.

\begin{figure}[tbp]
\begin{center}
  \includegraphics[width=0.35\columnwidth]{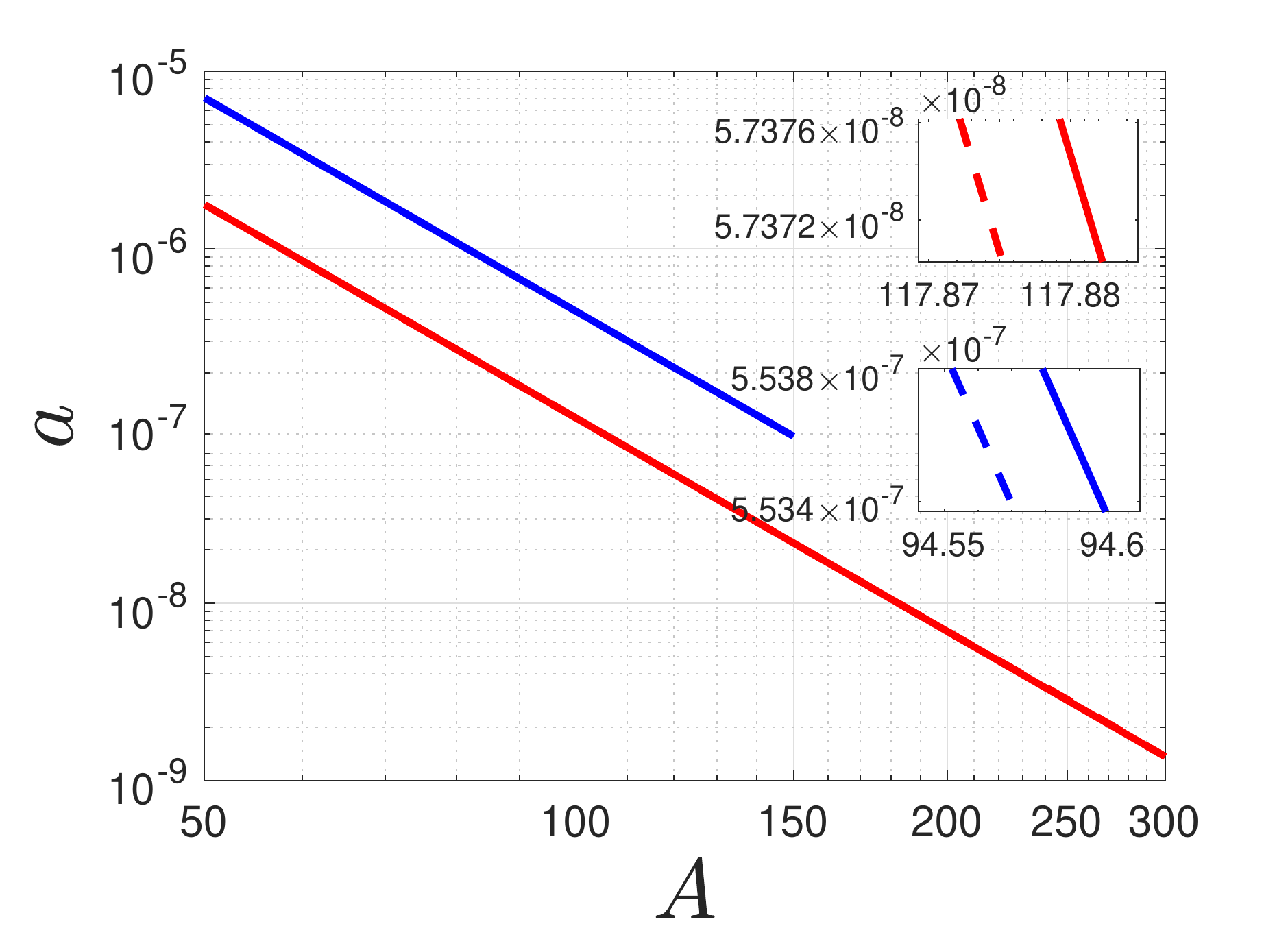}
  \includegraphics[width=0.35\columnwidth]{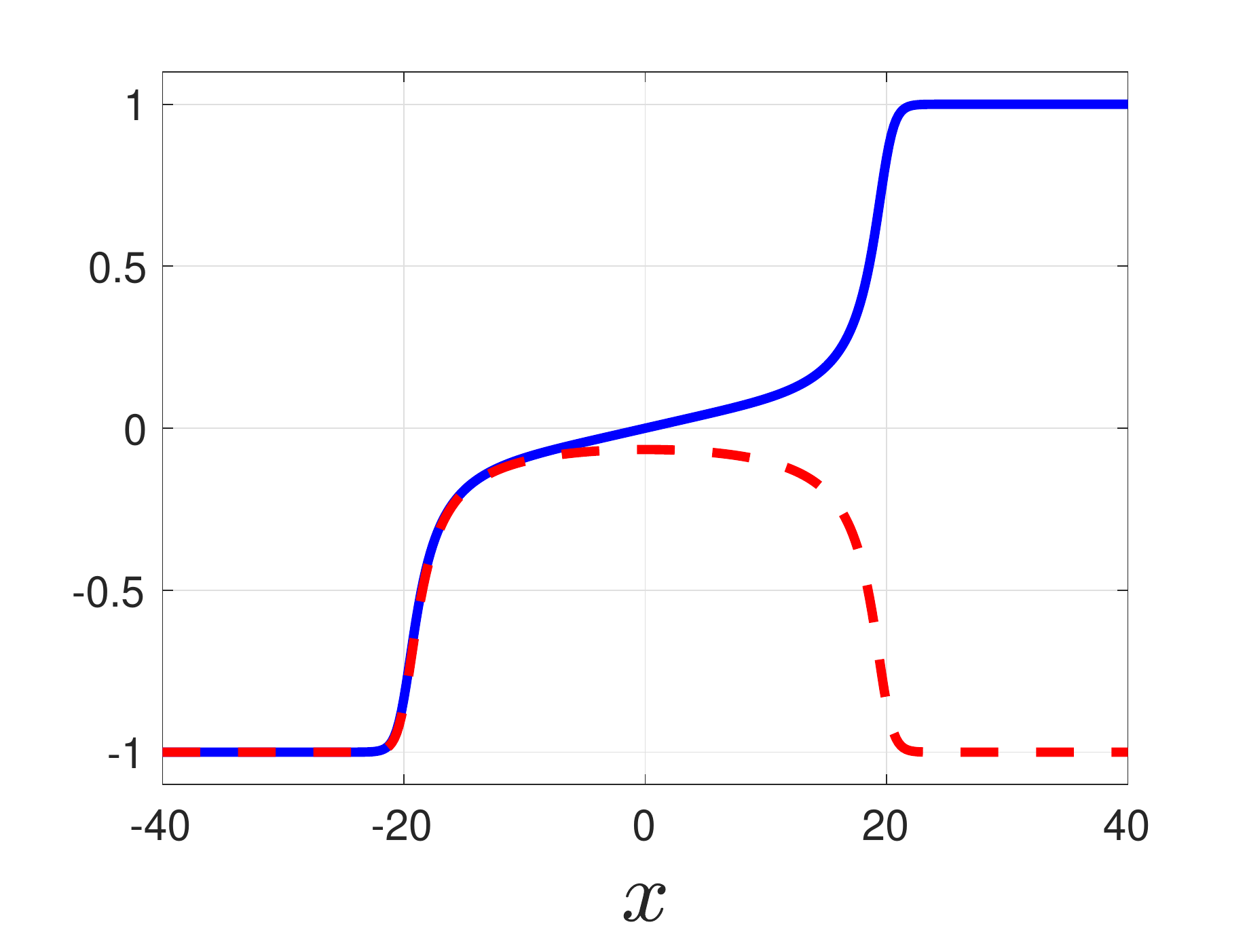}
  \includegraphics[width=0.35\columnwidth]{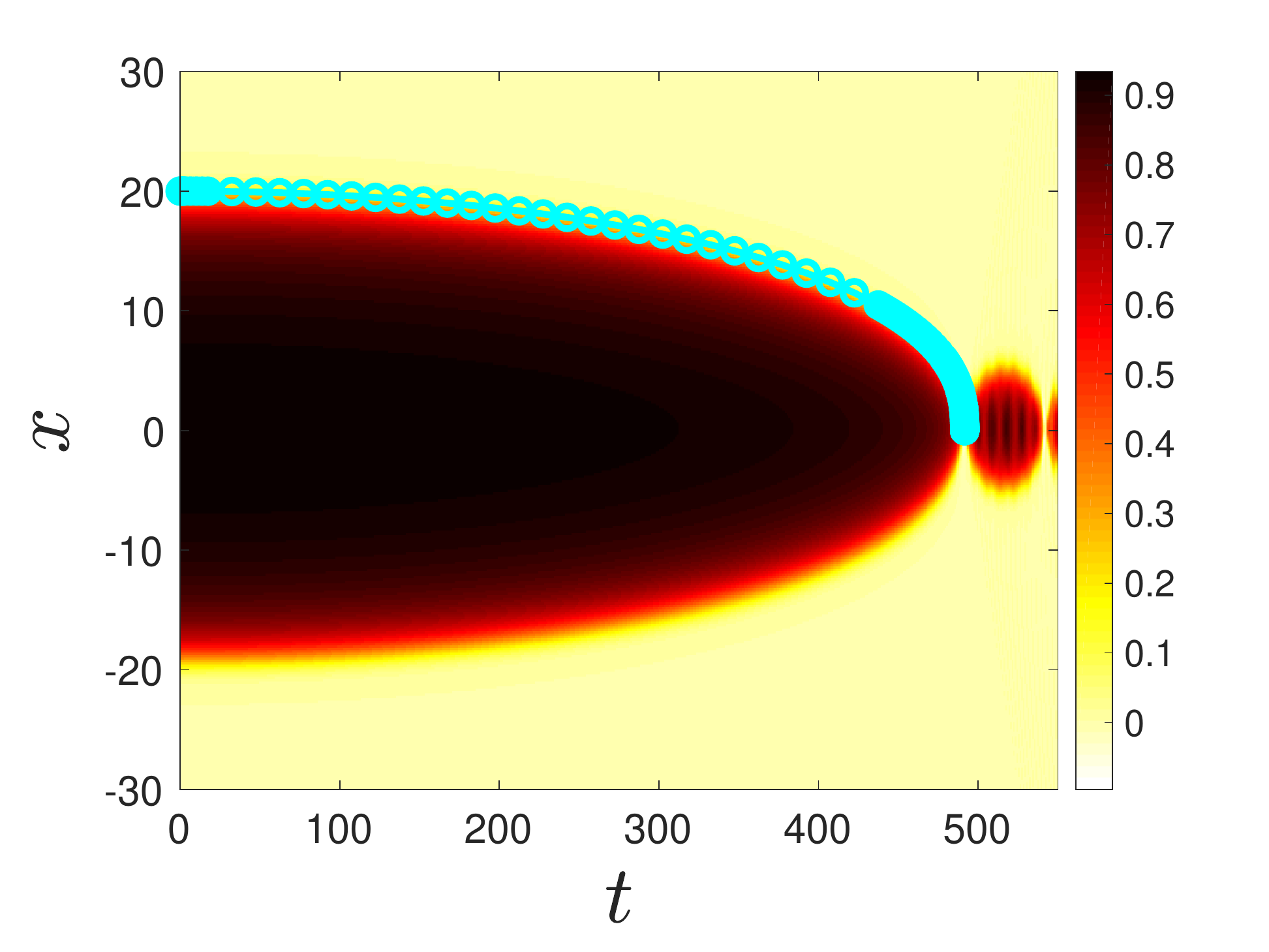}
  \includegraphics[width=0.35\columnwidth]{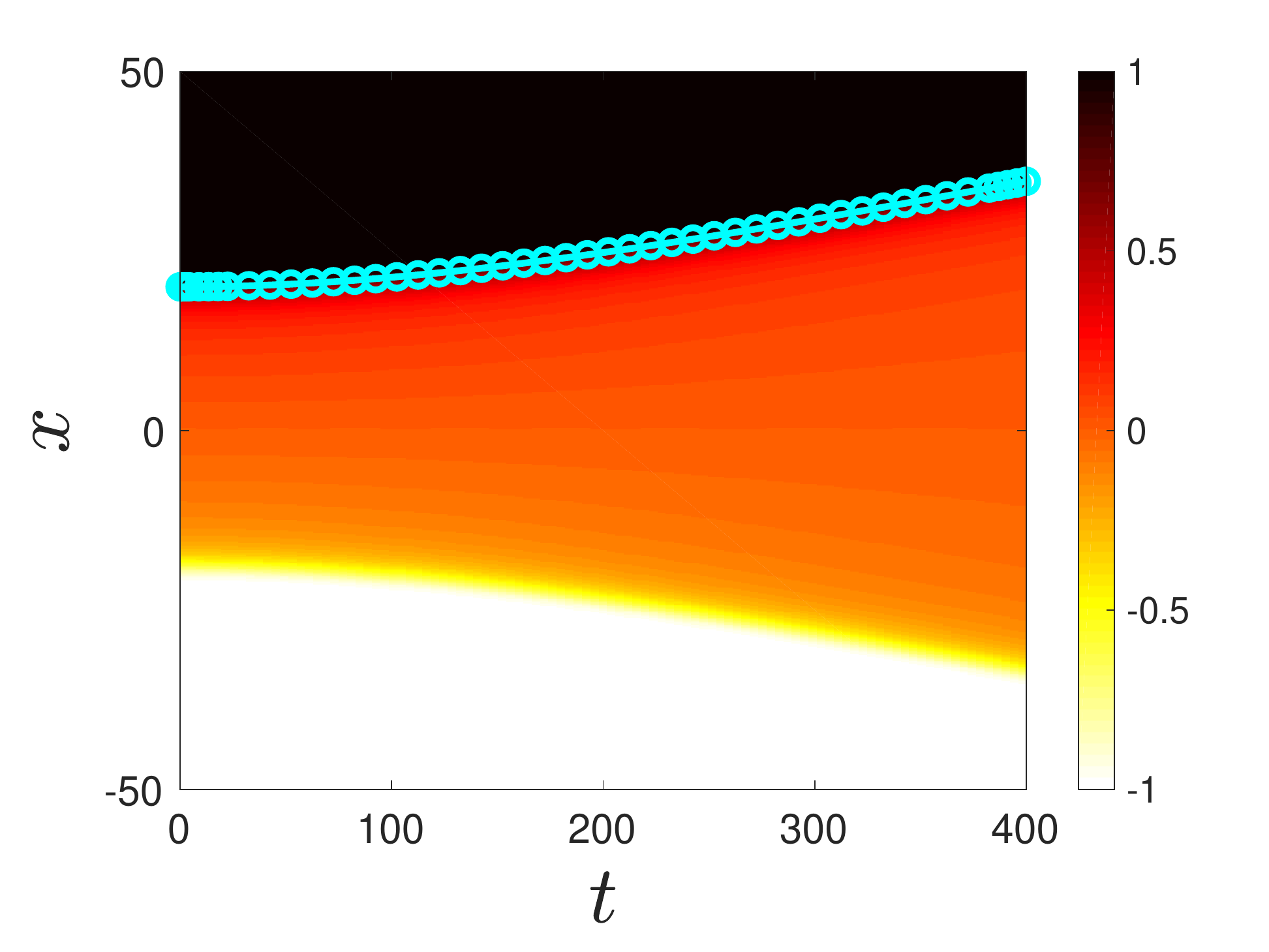}
  \caption{(Top left) Kink-antikink (red) and kink-kink (blue) acceleration
as a function of their separation. Inset shows the theoretical and numerical computations to be indistinguishable. (Top right) Typical
corresponding spatial profiles for the kink-antikink (dashed red) and the kink-kink (solid blue). (Bottom left) A kink-antikink simulation, displaying
attraction. (Bottom right) A kink-kink siumulation, displaying repulsion.
The circle symbols denote the results of the corresponding
one degree-of-freedom ODE reduction.
The relevant results have been obtained recently in~\cite{our}.
}
\label{fig6}
\end{center}
\end{figure}

Most recently, calculations in the past few
months have characterized the kink-antikink
and kink-kink asymptotic interactions in these higher order
models. The acceleration
of the kink-kink interaction as a function of their separation $X$
was found to be given by:
\begin{equation}\label{kink-kink}
a=\left[\frac{\Gamma(\frac{n-1}{2n})\Gamma(\frac{1}{2n})}{2n\sqrt{\pi}}\right]^{\frac{2n}{n-1}}\frac{(n+1)(n+3)}{4} X^{\frac{2n}{1-n}}.
\end{equation}
while  the acceleration of a kink-antikink pair is given by:
\begin{equation}\label{eqahat2a}
a=\left[\frac{-\sqrt{\pi}\,\Gamma(\frac{n-1}{2n})}{\Gamma(-\frac{1}{2n})}\right]^{\frac{2n}{{n-1}}}\frac{(n+1)(n+3)}{4} X^{\frac{2n}{1-n}}.
\end{equation}
These findings~\cite{manton2,our} were recently compared with numerical
computations yielding very good agreement both
for the kink-antikink and for the kink-kink interactions; see  Fig.~\ref{fig6}.
 
  \section{The Path Forward: Opportunities Ahead!}

  So, where does this all leave us? Integrable systems are, of course, 
  well understood. For non-integrable models, the ``flagship'' problem
  of the $\phi^4$ model still lays mathematically wide open!
  We have a good sense of what is going on---resonant transfer of energy from the propagating mode to an internal mode during the initial collision, and further exchange of the energy on later collisions lead to the fancy fractal multi-bounce windows. Yet, after 40 years, we currently have no satisfactory proof or quantitative characterization of the phenomenon.

  Then, there is a whole other world of Klein-Gordon models with higher-order nonlinearities. The next challenge is
  the $\phi^6$ model. It seems as if transient spectra and
  internal modes thereof are dynamically relevant. But it is unclear how to understand this in a mathematically convincing way.
  Another intriguing situation concerns systems that have not one, but two or many internal modes, as is controllably possible here. What happens to the resonance  picture in this setting?

  Higher order models offer another significant stepping stone.
  Here, the exponential decay (and its nice superposition properties
  due to the kinks' fast decay) are out the window. This has important
  implications for numerical simulations, where initial conditions must be chosen judiciously in order to avoid spurious effects.
  What are the implications of all this for the collisions and
  the resonance mechanisms? Are there internal modes in this
  case, and if so, what role do they play? Also, these systems possess asymmetric kinks with, on their two ends, exponential and  power law decay. Can we induce some sort of competition
  between a shorter distance but exponentially decaying 
  tail effect and a longer distance but fat (power law) tail
  effect? All these remain wide open questions in a theme that
  is just starting to warm up.

  It is amusing to think that kink collision numerical experiments
  started very early on in the history of solitonic dynamics. Yet,
  some four (and a half) decades later, they remain as challenging, as they
  appear (deceptively) simple.
  It seems that there are a few (or a lot) more kinks to iron out before the
  story is complete!

\bibliographystyle{abbrv}
\bibliography{references}

\begin{thebibliography}{10}

\bibitem{Steele2006}
{2006 Steele Prizes}.
\newblock {\em Notices Amer. Math. Soc}, 53:464, 2006.

\bibitem{ablowitz2}
M.~J. Ablowitz and P.~A. Clarkson.
\newblock {\em Solitons, Nonlinear Evolution Equations and Inverse Scattering}.
\newblock Cambridge University Press, Cambridge, 1991.

\bibitem{AKL}
M.~J. Ablowitz, M.~D. Kruskal, and J.~F. Ladik.
\newblock Solitary wave collision.
\newblock {\em SIAM J. Appl. Math.}, 36:428, 1979.

\bibitem{abl3}
M.~J. Ablowitz, B.~Prinari, and A.~D. Trubatch.
\newblock {\em Discrete and Continuous Nonlinear Schr{\"o}dinger Systems}.
\newblock Cambridge University Press, Cambridge, 2004.

\bibitem{ablowitz}
M.~J. Ablowitz and H.~Segur.
\newblock {\em Solitons and the Inverse Scattering Transform}.
\newblock SIAM, Philadelphia, 1981.

\bibitem{anni91}
P.~Anninos, S.~Oliviera, and R.~A. Matzner.
\newblock Fractal structure in the scalar $\lambda (\phi^{2}-1)^{2}$ theory.
\newblock {\em Phys. Rev. D}, 44:1147, 1991.

\bibitem{aub}
S.~Aubry.
\newblock {A unified approach to the interpretation of displacive and
  order--disorder systems. II. Displacive systems}.
\newblock {\em J. Chem. Phys.}, 64(3392), 1976.

\bibitem{kudr87}
T.~I. Belova and A.~E. Kudryavtsev.
\newblock Quasiperiodical orbits in the scalar classical $\lambda \varphi^4$
  field theory.
\newblock {\em Physics-Uspekhi}, 40:359, 1997.

\bibitem{Bondeson:1979bh}
A.~Bondeson, M.~Lisak, and D.~Anderson.
\newblock {Soliton Perturbations: A Variational Principle for the Soliton
  Parameters}.
\newblock {\em Physica Scripta}, 20:479--485, 1979.

\bibitem{sodano}
D.~K. Campbell, M.~Peyrard, and P.~Sodano.
\newblock Kink-antikink interactions in the double sine-gordon model.
\newblock {\em Physica D}, 19:165, 1986.

\bibitem{csw}
D.~K. Campbell, J.~F. Schonfeld, and C.~A. Wingate.
\newblock Resonance structure in kink-antikink interactions in $\phi^4$ theory.
\newblock {\em Physica D}, 9:1, 1983.

\bibitem{caputo1991}
J.~G. Caputo and N.~Flytzanis.
\newblock {Kink-antikink collisions in sine-Gordon and $\phi^4$ models:
  Problems in the variational approach}.
\newblock {\em Phys. Rev. A}, 44:6219, 1991.

\bibitem{caputo1994}
J.~G. Caputo, N.~Flytzanis, C.~Ragiadakos, and C.~Aignan.
\newblock {Removal of Singularities in The Collective Coordinate Description of
  Localised Solutions of Klein Gordon Models}.
\newblock {\em J. Phys. Soc. Jab.}, 63(7):2523--2531, 1994.

\bibitem{chong2012}
C.~Chong, D.~E. Pelinovsky, and G.~Schneider.
\newblock On the validity of the variational approximation in discrete
  nonlinear schr{\"o}dinger equations.
\newblock {\em Phys. D}, 241:115--124, 2012.

\bibitem{Christ:1975db}
N.~H. Christ and T.~D. Lee.
\newblock {Quantum expansion of soliton solutions}.
\newblock {\em Phys. Rev. D}, 12:1606--1627, 1975.

\bibitem{our}
I.~C. Christov, R.~J. Decker, A.~Demirkaya, V.~A. Gani, P.~G. Kevrekidis,
  A.~Khare, and A.~Saxena.
\newblock Kink-kink and kink-antikink interactions with long-range tails.
\newblock {\em Phys. Rev. Lett.}, 122:171601, 2019.

\bibitem{asli2}
I.~C. Christov, R.~J. Decker, A.~Demirkaya, V.~A. Gani, P.~G. Kevrekidis, and
  R.~V. Radomskiy.
\newblock Long-range interactions of kinks.
\newblock {\em Phys. Rev. D}, 99:016010, 2019.

\bibitem{oursg}
J.~Cuevas, P.~G. Kevrekidis, and F.~L. Williams, editors.
\newblock {\em The sine-Gordon Model and its Applications: From Pendula and
  Josephson Junctions to Gravity and High Energy Physics}.
\newblock Springer-Verlag, Heidelberg, 2014.

\bibitem{demirkaya}
A.~Demirkaya, R.~J. Decker, P.~G. Kevrekidis, I.~C. Christov, and A.~Saxena.
\newblock Kink dynamics in a parametric $\varphi^6$ system: a model with
  controllably many internal modes.
\newblock {\em J. High Ener. Phys.}, 2017:71, 2017.

\bibitem{celestial}
F.~Diacu and P.~Holmes.
\newblock {\em Celestial Encounters: The Origins of Chaos and Stability}.
\newblock Princeton University Press, Princeton, 1996.

\bibitem{gibbon}
R.~K. Dodd, J.~C. Eilbeck, J.~D. Gibbon, and H.~C. Morris.
\newblock {\em Solitons and Nonlinear Wave Equations}.
\newblock Academic Press, London, 1982.

\bibitem{dorey}
P.~Dorey, K.~Mersh, T.~Romanczukiewicz, and Y.~Shnir.
\newblock Kink-antikink collisions in the $\phi^6$ model.
\newblock {\em Phys. Rev. Lett.}, 107:091602, 2011.

\bibitem{drazin}
P.~G. Drazin and R.~S. Johnson.
\newblock {\em Solitons: An introduction}.
\newblock Cambridge University Press, Cambridge, 1989.

\bibitem{get}
B.~S. Getmanov.
\newblock New lorentz invariant systems with exact multisoliton solutions.
\newblock {\em {JETP} Lett.}, 24:291, 1976.

\bibitem{goodman2008}
R.~H. Goodman.
\newblock {Chaotic scattering in solitary wave interactions: A singular
  iterated-map description}.
\newblock {\em Chaos}, 18(2):023113, 2008.

\bibitem{good07}
R.~H. Goodman and R.~Haberman.
\newblock Kink-antikink collisions in the $\phi^4$ equation: The n-bounce
  resonance and the separatrix map.
\newblock {\em SIAM J. App. Dyn. Sys}, 4:1195, 2005.

\bibitem{goodman2005}
R.~H. Goodman and R.~Haberman.
\newblock {Kink-antikink collisions in the $\phi^4$ equation: The n-bounce
  resonance and the separatrix map}.
\newblock {\em SIAM J. Appl. Dyn. Sys.}, 4(4):1195--1228, 2005.

\bibitem{good15}
R.~H. Goodman and R.~Haberman.
\newblock Chaotic scattering and the n-bounce resonance in solitary-wave
  interactions.
\newblock {\em Phys. Rev. Lett.}, 98:104103, 2007.

\bibitem{ghw2002}
R.~H. Goodman, P.~J. Holmes, and M.~I. Weinstein.
\newblock {Interaction of sine-Gordon kinks with defects: phase space transport
  in a two-mode model}.
\newblock {\em Phys. D}, 161(1):21--44, Jan. 2002.

\bibitem{Kaup}
D.~J. Kaup and T.~Vogel.
\newblock {Quantitative measurement of variational approximations}.
\newblock {\em Phys. Lett. A}, 362(4):289--297, 2007.

\bibitem{ourp4}
P.~G. Kevrekidis and J.~Cuevas-Maraver, editors.
\newblock {\em A Dynamical Perspective on the $\phi^4$ model}.
\newblock Springer-Verlag, Heidelberg, 2019.

\bibitem{ivan}
A.~Khare, I.~C. Christov, and A.~Saxena.
\newblock {Successive phase transitions and kink solutions in
  ${\ensuremath{\phi}}^{8}$, ${\ensuremath{\phi}}^{10}$, and
  ${\ensuremath{\phi}}^{12}$ field theories}.
\newblock {\em Phys. Rev. E}, 90:023208, 2014.

\bibitem{kud}
A.~E. Kudryavtsev.
\newblock Solitonlike solutions for a {H}iggs scalar field.
\newblock {\em JETP Lett}, 22:82--83, 1975.

\bibitem{manton}
N.~S. Manton.
\newblock An effective {L}agrangian for solitons.
\newblock {\em Nucl. Phys. B}, 150:397--412, 1979.

\bibitem{manton2}
N.~S. Manton.
\newblock Forces between kinks and antikinks with long-range tails.
\newblock {\em J. Phys. A}, 52:065401, 2019.

\bibitem{peyrard}
M.~Peyrard and D.~K. Campbell.
\newblock Kink-antikink interactions in a modified sine-{G}ordon model.
\newblock {\em Physica D}, 33, 1983.

\bibitem{Scott2005}
A.~C. Scott.
\newblock personal communication, 2005.

\bibitem{sugi79}
T.~Sugiyama.
\newblock Kink-antikink collisions in the two-dimensional $\phi^4$ model.
\newblock {\em Prog. Theor. Phys.}, 61:550, 1979.

\bibitem{nls}
C.~Sulem and P.~L. Sulem.
\newblock {\em The Nonlinear Schr{\"o}dinger Equation}.
\newblock Springer-Verlag, New York, 1999.

\bibitem{wei3}
I.~Takyi.
\newblock Collective coordinate description of kink-antikink interaction.
\newblock Master's thesis, Stellenbosch University, 2016.

\bibitem{wei2}
I.~Takyi and H.~Weigel.
\newblock Collective coordinates in one-dimensional soliton models revisited.
\newblock {\em Phys. Rev. D}, 94:085008, 2016.

\bibitem{wei1}
H.~Weigel.
\newblock Kink--antikink scattering in $\phi^4$ and $\phi^6$ models.
\newblock In {\em J. Phys Conf. Series}, volume 482, page 012045, 2014.

\end{thebibliography}

\end{document}